# Quantum SI: The New System of Units

**Indulekha Kavila**
*School of Pure and Applied Physics, Mahatma Gandhi University*
*Kottayam 686 560, INDIA*
*Email: indulekha@mgu.ac.in, kindulekha@gmail.com*

**Abstract**

This article gives an overview of the new SI and a few details as to how its setting up was achieved. Those interested in the story of how the kilogram was fixed by tying it to the Planck constant can go straight to §5. Otherwise, apart from the introduction given as §1 which also presents excerpts from the relevant official documents on the key aspects of the new system, §2 gives a brief timeline of the history of the decimal metric system, §3 gives a few details of the procedural steps–from the setting up of a standard of measurement to its percolation to the public, taking the time standard as an example and, §4 attempts to give a flavour of the actual process of the setting up of a standard.

## §1 Introduction

Quite recently a few crystal balls entered our lives; quietly and without changing anything they made the everyday kilogram an avatar of the Planck constant of the quantum world! On 20[th] May 2019, the definition, of all seven basic measuring units of the International System of Units SI (abbreviated from the French *Système international (d'unités)*) in terms of seven invariant constants of nature, was formally accepted. With that the last remaining physical artefact in the SI system – a cylinder of metal known as the International Prototype of the Kilogram (IPK) – became obsolete. The kilogram got tied to the Planck constant instead, via balls shaped from $_{28}$Si single crystals. An artefact is something man-made as opposed to something natural.

Thus was realized the long standing goal expressed in the founding words of the decimal metric system when it took birth in the aftermath of the French Revolution–"for all peoples, for all times". The goal was to define and realize a system of base units of measurement that had long-term stability and would be accessible to all. Now, the natural constants have been given fixed values determined in the most accurate way possible, and the existing standard units have been reverse-defined using them. This article takes a brief look at the new SI, also called the quantum SI since the preferred constants are important in the theories of quantum mechanics. Below are given some of the key aspects of the definitions of the units in the new SI (see Boxes 1 and 2). Its crux is encapsulated in the SI logo (Figure 2).

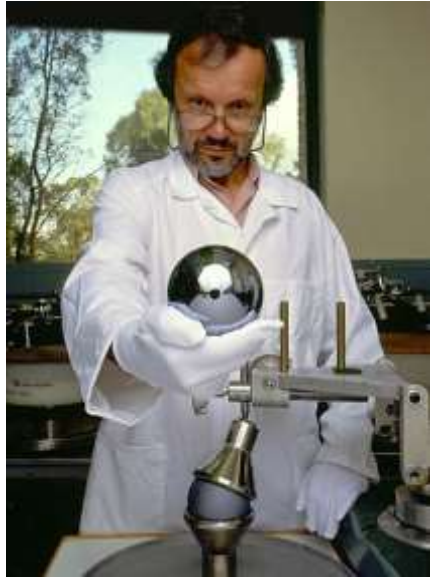

Figure 1: Achim Leistner at the Australian Centre for Precision Optics (ACPO) holds a 1 kg, single-crystal silicon sphere for the Avogadro project. Among the roundest man-made objects in the world, the sphere scaled to the size of Earth would have a high point of only 2.4 metres above "sea level". (*Figure courtesy*: The Commonwealth Scientific and Industrial Research Organisation of Australia, CC BY-SA 3.0, https://commons.wikimedia.org/w/index.php?curid=3517750)

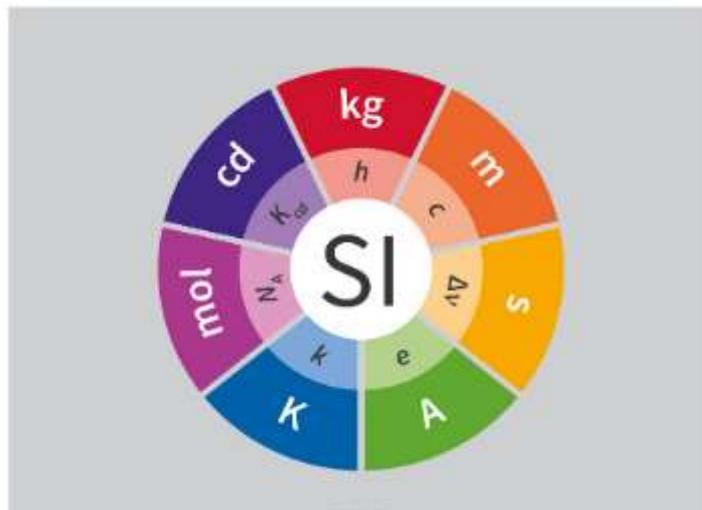

Figure 2: The new logo for the SI system
(*Figure*: courtesy of BIPM – (abbreviated from the French *Bureau International des Poids et Mesures*) International Bureau of Weights and Measures)

In general, in the new SI, any SI unit is a product of powers of some of these seven base units and a dimensionless factor. For example,

$$1 \text{ m} = \frac{9192631770}{299792458} \frac{c}{\Delta \nu_{Cs}}.$$

All quantities in use, other than the basic ones, are called "derived quantities" and are measured using derived units, which can be written as products of powers of the base units. For the list of twenty-two derived units which have been given special names see Table 2 of Ref [1]. A sample entry reads: "Quantity – electric potential difference; name of derived unit – volt; symbol for the unit – V; Expression in terms of base units – kg m$^2$s$^{-3}$A$^{-1}$= watt / ampere".

In the decimal metric system, the multiples / submultiples of the units are prescribed in terms of specific powers of ten. For the multiplying factors as well as the prefixes used for specifying each of the multiples / submultiples see Box 2 [1].

## §2 The long road to the quantum SI – some milestones

The journey towards better and finer ways of measurement, has now reached the new SI. The metric system was established in France on April 17, 1795. Over the last two hundred and twenty five years it has undergone revisions and augmentations to become the present SI. Some of the milestones along the long path, excerpted from the Press Kit of the 26[th] General Conference on Weights and Measures – 2018 [2] and adapted, is given in Box 3.

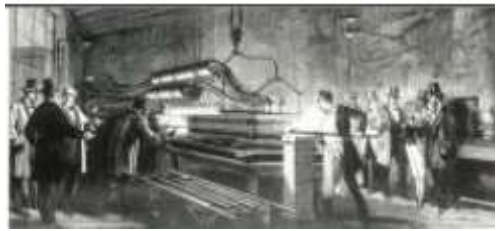

Figure 3: Casting of the platinum-iridium alloy used to manufacture national prototypes of the metre at the Conservatoire des Arts et Métiers in 1874. (*Photograph of engraving: courtesy of BIPM*)

## §3 The procedure

*Metrology*: Markets are known to have existed in the world from before 3000 BCE; market practices require regulation of weights and measures. Ancient Egyptians had used anthropic units of length, standardized using cubit rods, strands of rope, and official measures maintained at some temples. The system of time units detailed in various Indian *Puranas* go all the way from the *Truti* (~ microsecond), a fraction of the time required for batting an eyelid to the *Mahakalpa* (~$10^{22}$ seconds), the life span of *Brahma*. Every culture and every region had evolved its own range of units for measurement of mass, length and time! As societies became more and more connected, there arose the need for reliable and uniform measurements as well as units for additional measures, in trade and industry and later for scientific exchange. With the advancement of technology and the development of instruments functioning at higher and higher precision and also the development of space travel, the need for units that are universally reproducible and also flexible enough to accommodate future calls for greater precision, became imperative. The same philosophy which drove the evolution from grains of corn, the dimensions of the Pharaoh's arm, the batting of eyelids etc. to the metre as the length marked on the International Prototype Meter (a rod of Platinum (90%) and Iridium (10%) alloy) as the

10 millionth part of the quadrant of the earth passing through Paris, the kilogram as the mass of the International Prototype Kilogram (IPK – a cylinder made of an alloy of Platinum (90%) and Iridium (10%) with height (39 mm) approximately equal to diameter and) as the thousandth part of the mass of a cubic meter of distilled water and the second as the eighty six thousand four hundredth part of a mean solar day, led to the taking of the steps that now link the basic measuring units to fundamental constants of nature.

Metrology is a dynamic branch of science – '*studied by scientists and improved by engineers*'. It is the science of measurement and its application, including all '*theoretical and practical aspects of measurement, irrespective of measurement uncertainty and the particular field of application*'. Metrologists work on defining basic units that are in conformity with the stringent requirements of usage in various fields, reliability, precision, stability, reproducibility, accessibility, etc. Also, they develop procedures for percolating these units to the public domain for general use.

*The procedure for the second*: Time units, fixed using Earth's rotation period as the base and the second as 1 / 86400 of a mean solar day had been in use long. This was replaced with a new definition for the second, based on Earth's orbital motion, since the solar day was found to be getting longer as well as fluctuating. The variations are due respectively, to torquing by the Moon and to changes in moment of inertia arising from the coupling between Earth's core and mantle. The International Committee on Weights and Measures defined the second as 1/31,556,925.9747 of the length of the tropical (seasonal) year 1900. The definition was ratified in 1960. This measure called the ephemeris second, based on the fixedness of Earth's rotational angular momentum had also to be discarded soon. By the 1960's, with the development of atomic clocks, precision time keeping to an accuracy of 2 parts in $10^{10}$ became possible. A worldwide campaign to determine exactly the frequency of the hyperfine transition of Cs 133 gave results that varied by $\pm 30$ Hz. The unexpected variations were traced to variations in the length of the year and consequent variations in the ephemeris second defined as a fixed fraction of the year. Torqueing by planets causes variations in the orbital period. Thus, advances in molecular beam spectroscopy which led to the development of atomic clocks, pointed out an urgent need for a stable time standard also. Below we give relevant excerpts from the *mise en pratique*, prepared by the Consultative Committee for Time and Frequency (CCTF) of the International Committee for Weights and Measures–CIPM (abbreviated from the French *Comité international des poids et mesures*), to indicate how the new definition of the SI base unit, the second, symbol s, got realized and was put into practice [3]. *Mise en pratique* is a French phrase meaning "to put in practice", and by way of application a *mise en pratique* is a sequence of steps that are enacted to accomplish a desired outcome. The details given below are an adapted and curtailed version of the *mise en pratique* in respect of the unit for time given in Ref [3].

First we give the meaning of some of the phrases that will be used. The term "to realize a unit" is interpreted to mean the establishment of the value and associated uncertainty of a quantity of the same kind as the unit that is consistent with the definition of the unit. Thus for example, the definition of the second stands independent of any particular experiment designed for realizing it in practice. In fact, any method capable of deriving a time value that can be expressed in terms of any or all of the set of seven reference constants could, in principle, be used. For each unit, the relevant Consultative Committee gives a list of methods that are easiest to implement and/or that provide the smallest uncertainties, and as such are officially recognized as primary methods. A primary method is a method having the highest metrological properties; whose operation can be completely described and understood; for which a complete uncertainty statement can be written down in terms of SI units; and which does not require a reference standard of the same quantity.

1. *The definition*–Example: The definition of the second, SI base unit of time, is as follows–The second, symbol s, is the SI unit of time. It is defined by taking the fixed numerical value of the caesium frequency $\Delta\nu_{Cs}$, the unperturbed ground-state hyperfine transition frequency of the caesium 133 atom, to be 9 192 631 770 when expressed in the unit Hz, which is equal to $s^{-1}$.

2. *The practical realization*–Example: Atomic clocks offer a means for the practical realization of the second. The definition of a unit refers to an idealized situation. Thus, the definition of the second has to be understood as referring to atoms free of any perturbation, at rest and in the absence of electric and magnetic fields. Also, since there will be time dilation effects due to the Earth's gravitational field as per the General Theory of Relativity, the second should be understood as the definition of the unit of proper time: it applies in a small spatial domain which shares the motion of the caesium atom used to realize the definition. Thus, in a laboratory sufficiently small to allow the effects of the non-uniformity of the gravitational field to be neglected when compared to the uncertainties of the realization of the second, the proper second can be obtained after application of a) the special relativistic correction for the motion of the atom b) the black body radiation shift that will take into account the change in the energy levels due to the electric component of the permeating general radiation field (Stark splitting) – the magnetic part of this field being negligible compared to that of the electric field – and estimates of the corrections for several other effects related to the clock design and operation. In practice, the realization is designed and built to capture the defined quantity with the least possible uncertainty that can be experimentally attained.

3. *Primary standards*–A small number of national metrology laboratories realize the particular unit with the highest accuracy. To do so, they design and build standards that reproduce the relevant measure with a very low uncertainty after all relevant corrections have been estimated and applied. In the case of the second, by 2017, the best of these primary standards could produce the SI second with a relative standard uncertainty almost approaching one part in $10^{16}$. It may be noted that at such levels of accuracy the general relativistic corrections for the change in the gravitational field across the device cannot be ignored. The standard will thus have to be calibrated using general relativity to provide the proper time at a specified point in the device.

4. *Secondary representations*–A list of secondary representations which can realize the particular unit with greater accuracy but only with uncertainties that are much higher since they are currently linked to the current primary standard, are also maintained by the BIPM.

5. *Other standards*–These could be commercially available devices whose accuracy compare well with the primary standard and could be used as references for fixing the particular unit by national laboratories that act as keepers of the standard.

6. *For public use*–measuring devices may be calibrated by laboratories that act as keepers of the standard.

The connection between time elapsed as per the standard, to the time in common use is discussed below. To keep track of time elapsed, the need to compare and synchronize clocks situated in widely separated establishments arise. In the context of general relativity, the concept of synchronization is arbitrary, and is done on the basis of a convention for simultaneity and synchronization. The Global Navigation Satellite Systems (GNSS) provide a satisfactory solution to the problem of time transfer. It may be noted that in addition to the TAI and the UTC discussed below, there are various times defined for astronomical purposes (see for example the International Earth Rotation and Reference Systems Service conventions).

A continuous time scale called International Atomic Time (TAI), is produced by the BIPM based on the best realizations of the SI second. A set of atomic clocks throughout the world (operated by National laboratories) keeps time by consensus. That is, the clocks "vote" on the correct time and a consensus is taken. The time arrived at by consensus is called International Atomic Time (TAI). Then, all voting clocks are steered to agree with the TAI. The whole exercise is repeated regularly. TAI "ticks" atomic seconds and is more stable and more accurate than any of the individual contributing clocks.

TAI is not distributed directly in everyday life. The time in common use is Coordinated Universal Time (UTC). UTC is a time scale produced by the BIPM with the same rate as TAI, but differing from TAI only by an integral number of seconds, by definition. This difference will be modified in steps of 1 s, using a positive or negative leap second (similar to the use of an extra day in leap years for adjusting calendar years against the orbital period of the Earth), in order to keep UTC in agreement with the time defined by the rotation of the Earth such that, when averaged over a year, the Sun crosses the Greenwich meridian at noon UTC to within 0.9 s. UTC "ticks" the same atomic seconds as TAI, but inserts or omits leap seconds as needed, to correct for variations in the rate of rotation of the earth.

National time-service laboratories, which contribute to the formation of UTC by the BIPM, maintain an approximation of UTC, known as UTC(k) for laboratory k. Laboratories are to maintain the local realizations UTC(k) within 100 ns offset from UTC. In fact, the clocks kept by 35 of the 78 laboratories are accurate enough that their offsets amount to only a few tenths of nanoseconds. In some cases, UTC(k) represents the basis of the legal time in the respective country. For example Indian Standard Time is maintained by CSIR – National Physical Laboratory which is India's National Measurement Institute. Legal times are set such that they are in general offset from UTC by an integer whole (with exceptions) number of hours to establish time zones and daylight-saving time. Such legal times are disseminated by various means, depending on the country, such as dedicated time-signal transmitters, radio, television, the speaking clock, telephone lines, the Internet, and dedicated fibre-optic transmission services. In addition, each of the GNSS de facto serves as a means for disseminating a prediction of UTC, with deviations from UTC by a few ten nanoseconds or better.

### §4 The process

The new SI uses the rules of nature, which fix the values of natural constants, to create the rules of measurement for fixing the base units. Thus, the latest SI links measurements at the finest levels that are accessible to us now, viz. the atomic and quantum scales, to those at the macroscopic level. (It may be remembered that the constancy of the fundamental constants of nature, are checked through various cleverly set up astronomical observations, across the vastness of astronomical time scales [4].) The new definitions, which reproduce the old measures as closely as possible, are then used with accuracies that are in compliance with the needs of economy, science and society.

The carrier of the definition of the SI unit and the carriers of the realization of the SI unit are separate; the particular natural constant is the carrier of the definition and particular standards that are set up are the carriers of the realizations. The process of realizing a given primary standard for any given unit is complex and multifaceted, involving large numbers of setups and groups and, scientists and engineers. Take the counting of the number of wave crests in an electromagnetic wave – specifically to the tune of 9 192 631 770 per second needed for setting up the time standard. The particular line from the Cs-133 standard, with a frequency ~ 9.1 GHz, falls in the microwave region. The counting can be done using frequency counters. (Cs-133 is the lone stable isotope of Caesium; presence of isotopes can reduce the sharpness of the primary line of interest since the isotopes will produce lines that have a slightly different central frequency). In high frequency counters, like those for microwaves, a prescaler scales the frequency down to a point where normal digital circuitry can do the counting. In the case of higher frequencies the signal is combined with another periodic signal of known frequency to get a signal at the difference frequency which is low enough to be counted directly (Sin A Cos B = $\frac{1}{2}(Sin\,(A+B) + Sin\,(A-B))$!).

In frequency counters a counter accumulates the number of events occurring within a specific period of time after which the accumulated value is transferred to a display and the counter reset to zero. An internal oscillator called the timebase keeps track of time. Highly accurate electronic circuits, which use quartz crystal oscillators sitting in highly controlled ambient conditions, are used to generate timebases. Such frequency standards generate a fundamental frequency with high precision; also, the harmonics provide reference points. A time standard may be derived from the frequency standard since time is the reciprocal of frequency.

The core of an atomic clock is a tunable microwave cavity. In a caesium atomic clock, Caesium atoms are prepared in a hyperfine state and a beam of these move through a cavity. The atoms of the gas when irradiated with microwaves of the appropriate frequency absorbs the microwaves. The cavity contains an electronic amplifier to make it oscillate. The number of atoms present in the chamber at any time, that have changed state / remains unchanged, is detected by irradiation with another laser and tracking the subsequent fluorescence for example. Either way, the number of atoms which change hyperfine state is detected and the cavity is tuned for a maximum of detected state changes. This maximum happens, when the gas is irradiated with waves at the same frequency as that of the hyperfine transition. Clock adjustment is rendered complex due to the need to correct for unwanted side-effects, such as frequencies from other electron transitions, temperature changes, and the spreading in frequencies caused by the fact that the emission / absorption is by an ensemble of atoms. The velocities of the atoms will be slightly different from each other; this introduces Doppler shifts that differ between the emissions from different atoms. To control long-term drifts in the frequency of the system, the microwave oscillator's frequency is swept across a narrow range. A modulated signal will be produced at the detector. The detector's signal can then be demodulated to apply negative feedback. In this way, the quantum-mechanical properties of the atomic transition frequency of the caesium can be used to tune the microwave oscillator to the same frequency, except for a small amount of experimental error. And the particular microwave frequency may be counted using a frequency counter with a timebase that has been calibrated against the ephemeris second.

Lord Kelvin had suggested the use of atomic transitions to measure time in 1879. A practical method for doing this came up when Isidor Rabi developed experimental set ups that achieved magnetic resonance in beams of particles in the 1930s. In fact, Rabi himself had suggested that atomic beam magnetic resonance might be used as the basis of such a clock.

The first accurate atomic clock, based on the caesium standard was built by Louis Essen and Jack Parry in 1955 at the National Physical Laboratory in the UK. In 1967 the calibration of the caesium standard atomic clock against the ephemeris second was carried out. Atomic clocks can keep time with an accuracy up to 1 part in $10^{14}$; equality of the ephemeris second (which is known with much less accuracy) with the SI second has been verified to within 1 part in $10^{10}$.

*Increasing the precision*: Lowering the temperature of the sample atoms helps. Colder atoms move much more slowly, allowing longer probe times. Spectral lines at higher frequencies and narrower lines allow better precision. Advances in instrumentation, like laser cooling and trapping of atoms, development of high-finesse Fabry-Perot cavities for narrower laser line widths, precision laser spectroscopy, optical frequency combs (OFC) which have made the counting of optical frequencies easier and simpler (before OFCs, which are actually mode locked lasers, it required about 10 scientists, 20 different oscillators and 50 feedback loops to perform a single optical measurement!), have all contributed to reducing the uncertainties in the measurement of time.

For use in defining the new units, there was a need for determining the numerical values of $h$, $e$, $k$ and $N_A$ to high precision. The Committee on Data for Science and Technology (CODATA), had fixed the numerical values of these constants by making, for each constant, a special least-squares adjustment to the values determined by different dedicated groups each of which had determined the constant by a different method (CODATA set of recommended values may be found on the World Wide Web at codata.org or at physics.nist.gov/constants). Since 2011, the International Avogadro Coordination had measured the Avogadro constant by counting the atoms in enriched $_{28}$Si monocrystals. With the determination of $N_A$, the mole was incorporated in the SI, removing the uncertainty between 'amount of substance' and 'count of entities'. The step is useful in the context of ultra-low level precision measurements that have become necessary since chemistry as well as biology are now considering increasingly smaller domains involving fewer and fewer molecules / atoms / ions / electrons etc. In the new SI, the Joule will replace the kelvin as a base unit. This can be effected by fixing the value of the Boltzmann constant $k$ which is the proportionality constant between temperature and thermal energy $k$T. The titles of three papers from the temperature community, reporting on the determination of the Boltzmann constant speak their own tale–"*New measurement of the Boltzmann constant k by acoustic thermometry of helium-4 gas*"; "*Final determination of the Boltzmann constant by dielectric-constant gas thermometry*"; "*An improved electronic determination of the Boltzmann constant by Johnson noise thermometry*".

## §5 Enter balls of crystalline $^{28}$Si

(*supported by institutions like the BIPM, NIST (National Institute of Standards and Technology, USA), INRIM (Istituto Nazionale di Ricerca Metrologica, Italy), PTB (Physikalisch Technische Bundesanstalt, Germany) and NMIJ (National Metrology Institute of Japan); coordinated by IAC (International Avogadro Coordination)*)

The IPK was in service from 1889. Comparison with its official copies showed that the copies were increasing in mass at an average rate of about 0.5 μg per century. Any changes that are systematic and are common to the IPK and the copies would not be evident in any comparisons amongst themselves either. In effect, the numerical value of $m_p$, the mass of the proton in kg, based on the IPK, would drift! In view of all this, from about 1950, efforts aimed at anchoring the kilogram to the constants of nature were started by mass metrologists. The aim was to tie the kilogram to the Planck constant; $h$ has dimensions $[M][L]^2[T]^{-1}$. But, $h$ was difficult to measure and was known only with a relative uncertainty of ~ 50 x $10^{-9}$. Balls of crystalline $^{28}$Si enter the scene as part of the efforts aimed at determining $h$ to a sufficiently high accuracy. The fruition of these efforts culminated with the formal adoption of the new definitions of the SI units on May 20 2019. Behind the announcement lies decades of intense effort. The following is intended to capture a rather diluted flavour of the efforts towards the redefinition of the kilogram.

*Disconnecting the kilogram from the IPK*: The new definition of the kilogram rests on the following facts: a) the density of a perfect crystal is the same from the level of the unit cell to the bulk, the small mass difference arising due to the binding energy of the crystal and the thermal energy of the atoms in the crystal being negligibly small for the purpose at hand; i.e. for a perfect $^{28}$Si crystal, which has 8 atoms in a unit cell, Mass of crystal / Volume of crystal = 8 times mass of $^{28}$Si atom / volume of unit cell – linking macroscopic mass to atomic mass b) quantum electrical measurements can be made with very high precision (*see Box 4*) c) lattice constants may be determined with high precision by the X-Ray Crystal Density Method (XRCD) d) the mass of a macroscopic object can be measured through electrical means via the watt balance (*see Box 5*), linking macroscopic mass to electrical quantities and the already fixed length and time standards.

The facility offered by the facts listed above was realized through a new artefact–a crystal ball, a 1 kg sphere shaped out of a perfect crystal of $^{28}$Si. Any crystal would do. But silicon was chosen because the semiconductor industry had developed methods for growing large single crystals of silicon with extreme chemical purity and no dislocations. The absence of dislocations permits the determination of the lattice constant to high accuracy by the XRCD method.

Suppose we had a 100% perfect crystal of $^{28}$Si. Let its mass determined by the watt balance (also known as Kibble balance in honour of the scientist Bryan Kibble who proposed and implemented the concept) be $m_{artefact}$ (traceable to the IPK) and volume be determined as $V_{artefact}$. Let its lattice constant determined by the XRCD method (X-Ray Crystal Density method – an X-ray interferometric method) be *a*. The unit cell is cubic and each unit cell has 8 atoms. Then the total number of $^{28}$Si atoms in the artefact $N = 8\, V_{artefact}\,/\,a^3$. Therefore, amount of substance in the artefact $n = N / N_A = 8\, V_{artefact}\,/\,N_A\,a^3$ moles. Also, $N = m_{artefact}\,/\,m(^{28}\text{Si})$ and n = $m_{artefact}\,/\,M(^{28}\text{Si})$ moles, where $M(^{28}\text{Si})$ is the molar mass of $^{28}$Si. Relative atomic / electron masses $A_r(X) = m(X)\,/(^{12}\text{C}) = M(X)/M(^{12}\text{C})$ can be determined to high accuracy. Then, expressed in terms of relative mass $n = m_{artefact}\,/\,M(^{12}\text{C})A_r(^{28}\text{Si})$. Equating the two expressions for *n*, $N_A$ may be isolated as $N_A = 8\, V_{artefact}\, M(^{12}\text{C})A_r(^{28}\text{Si})\,/\,m_{artefact}\,a^3$ yielding

$$N_A / M(^{12}\text{C}) = 8\, V_{artefact}\, A_r(^{28}\text{Si})\,/\,m_{artefact}\,a^3 \qquad (1).$$

Now, the measured mass $m_{artefact}$ shall be expressed in terms of the parameters that are actually measured in the watt balance. The measurement using the watt balance can be traced to the IPK. The measurements yield the relation $m_{artefact}\,gv = U_m\,U_w\,/R_w$, where the current $I_w$ has been expressed via accurately measured quantities–the potential drop across a given resistance. Now *h* enters as follows [8]. Converting from standard units for voltage and resistance to their SI units involve multiplication with ($K_J$(standard units) / $K_J$) and ($R_K$/$R_K$(standard units)) respectively. By the Josephson and quantum-Hall-effect relations, the Josephson constant $K_J$ and the von Klitzing constant $R_K$ are given in terms of fundamental constants as $K_J = 2e/h$ and $R_K = h/e^2$. Substituting,

$$m_{artefact} g v = (h/4)(U_m U_w / R_w)[K_J \text{(standard units)}]^2 R_K \text{(standard units) Joule s}^{-1}$$

with $K_J$(standard units)] and $R_K$(standard units)] known exact values. Reordering terms we get

$$h = 4\, m_{artefact} g v / (U_m U_w / R_w) [K_J \text{(standard units)}]^2 R_K \text{(standard units)} \qquad (2),$$

relating the Planck constant to a macroscopic mass; here, other than via $m_{artefact}$ the mass dimension enters only via $h$.

The right hand sides of equations 1 and 2 involve measured quantities pertaining to the artefact. Now, consider the product of the left hand sides of equations 1 and 2. Using the relation between the Planck, Rydberg and fine structure constants viz. $h = \alpha^2 c m_e / 2R_\infty$, we get LHS1 x LHS2 = $h N_A / M(^{12}C) = \alpha^2 c (\frac{N_A m_e}{M(^{12}C)})/2R_\infty = \alpha^2 c A_r(e) /2R_\infty$. Thus the product of the terms on the LHS of the two equations can be calculated to high accuracy, independent of the watt balance or XRCD methods. Equations 1 and 2 were written assuming an ideal crystal. Since, in reality a 100% perfect, monoisotopic and impurity-free crystal of $^{28}$Si is practically impossible, there will be a mismatch with the value of the product RHS1 x RHS2 (which does not involve $m_{artefact}$!) evaluated from the indicated measured quantities. The value of this product, will not be sufficiently close to the calculated value of $\alpha^2 c A_r(e) /2R_\infty$. The closeness of the products LHS1 x LHS2 and RHS1 x RHS2 quantifies the perfection of the crystal ball. The crystal ball may be refined to bring the values of the two products as close as needed for achieving the desired level of precision. Once that is achieved RHS2 will evaluate to the value of $h$ to the desired level of accuracy.

The success of the collaborative efforts of the various institutions in achieving the feat lies in the fact that, the relative fraction of $^{28}$Si in the crystal ball could be made sufficiently high and the fraction and types of various impurity atoms and vacancies could be sufficiently accurately quantified. The whole process involved a very large number of steps broadly intended for a) $^{28}$Si isotope enrichment (by centrifugation of silicon tetrafluoride gas through cascades of hundreds of centrifuges, removal of chemical impurities by float-zone purification etc.) b) production of a crystal without dislocations (crystal grown using a crucible-free pedestal method) c) perfection of shape of the sphere cut out from $^{28}$Si single crystals (sphere has minimum surface area for a given volume and the surface is smooth without corners) d) characterization and quantification of the point defects in the crystal (mainly Carbon, Oxygen, Boron, etc. atoms and vacancies via infrared absorption spectroscopy, deep level transient spectroscopy, glow discharge mass spectrometry etc.) e) estimation of the adsorbed water in the surface layers (like by degassing silicon wafers etc.) f) quantification of the thickness and density of the surface layer which can be carbonaceous as well as have the oxide present (like via spectroscopic ellipsometry, x-ray photoelectron spectroscopy etc. ) etc.. Necessary modifications to equations (1) and (2) were made, to incorporate the effects of all the various types of imperfections and repeated and better attempts were made till a match could be obtained *(for some more details see Box 6)*. Thus the Avogadro constant $N_A$ as well as the Planck constant $h$ could both be estimated to high accuracy. This permitted the fixing of the value of $h$ with an accuracy that was sufficiently high to permit the reverse defining of the kilogram. Equation 2, may be looked at now as an equation for the mass, in terms of $h$, at the level of precision demanded by modern methods of measurement.

The Global Positioning System (GPS) requires time to be specified to nanosecond precision. Long before the GPS came into common use, metrologists had redefined the second so that time intervals could be measured to such precision. Whether in ever further refined experiments or in daily life, metrology permits the full use of technological advancements even as they come up, by defining units that measure up. Metrologists deliver, even if they have to look into crystal balls and count the atoms inside to do it!

**Acknowledgements**

I acknowledge material and tables from various documents put up on the internet by the BIPM that I have referenced below [1, 2, 3, 8]. I also thank *Wikimedia commons*, BIPM and NIST as the sources for photographs.

Box 1.  The seven base units of the SI

The definition of the new SI unit is as follows [1].  The SI is the system of units in which:
- the unperturbed ground state hyperfine transition frequency of the caesium 133 atom $\Delta v_{Cs}$ is 9 192 631 770 Hz,
- the speed of light in vacuum $c$ is 299 792 458 m/s,
- the Planck constant $h$ is 6.626 070 15 × $10^{-34}$ J s,
- the elementary charge $e$ is 1.602 176 634 × $10^{-19}$ C,
- the Boltzmann constant $k$ is 1.380 649 × $10^{-23}$ J/K,
- the Avogadro constant $N_A$ is 6.022 140 76 × $10^{23}$ mol−1,
- the luminous efficacy of monochromatic radiation of frequency 540 × $10^{12}$ Hz $K_{cd}$ is 683 lm/W,

where the hertz, joule, coulomb, lumen, and watt, with unit symbols Hz, J, C, lm, and W, respectively, are related to the units second, metre, kilogram, ampere, kelvin, mole, and candela, with unit symbols s, m, kg, A, K, mol, and cd, respectively, according to Hz = $s^{-1}$, J = kg $m^2$ $s^{-2}$, C = A s, lm = cd $m^2$ $m^{-2}$ = cd sr, and W = kg $m^2$ $s^{-3}$. These definitions specify the exact numerical value of each constant when its value is expressed in the corresponding SI unit (*see Box 1*).  By fixing the exact numerical value the unit becomes defined, since the product of the numerical value and the unit has to equal the value of the constant, which is invariant.

| Quantity | SI unit |
|---|---|
| time | The second, symbol s, is the SI unit of time. It is defined by taking the fixed numerical value of the caesium frequency $\Delta v_{Cs}$, the unperturbed ground-state hyperfine transition frequency of the caesium 133 atom a caesium atom at rest at a temperature of absolute zero, to be 9 192 631 770 when expressed in the unit Hz, which is equal to $s^{-1}$. |
| length | The metre, symbol m, is the SI unit of length. It is defined by taking the fixed numerical value of the speed of light in vacuum $c$ to be 299 792 458 when expressed in the unit m $s^{-1}$, where the second is defined in terms of $\Delta v_{Cs}$. |
| mass | The kilogram, symbol kg, is the SI unit of mass. It is defined by taking the fixed numerical value of the Planck constant $h$ to be 6.626 070 15 ×$10^{-34}$ when expressed in the unit J s, which is equal to kg $m^2$ $s^{-1}$, where the metre and the second are defined in terms of $c$ and $\Delta v_{Cs}$. |
| electric current | The ampere, symbol A, is the SI unit of electric current. It is defined by taking the fixed numerical value of the elementary charge $e$ to be 1.602 176 634 ×$10^{-19}$ when expressed in the unit C, which is equal to A s, where the second is defined in terms of $\Delta v_{Cs}$. |
| thermodynamic temperature | The kelvin, symbol K, is the SI unit of thermodynamic temperature. It is defined by taking the fixed numerical value of the Boltzmann constant $k$ to be 1.380 649 ×$10^{-23}$ when expressed in the unit J $K^{-1}$, which is equal to kg $m^2$ $s^{-2}$ $K^{-1}$, where the kilogram, metre and second are defined in terms of $h$, $c$ and $\Delta v_{Cs}$. |
| amount of substance | The mole, symbol mol, is the SI unit of amount of substance. One mole contains exactly 6.022 140 76 × $10^{23}$ elementary entities. This number is the fixed numerical value of the Avogadro constant, $N_A$, when expressed in the unit $mol^{-1}$ and is called the Avogadro number. The amount of substance, symbol n, of a system is a measure of the number of specified elementary entities. An elementary entity may be an atom, a molecule, an ion, an electron, any other particle or specified group of particles. |
| luminous intensity | The candela, symbol cd, is the SI unit of luminous intensity in a given direction. It is defined by taking the fixed numerical value of the luminous efficacy[1] of monochromatic radiation of frequency 540 ×$10^{12}$ Hz, $K_{cd}$, to be 683 when expressed in the unit lm $W^{-1}$, which is equal to cd sr $W^{-1}$, or cd sr $kg^{-1}$ $m^{-2}$ $s^3$, where the kilogram, metre and second are defined in terms of h, c and $\Delta v_{Cs}$. |

[Table courtesy: Ref [1]]

Margin Note 1

Luminous flux means, that part (here the visible part) of the flux of electromagnetic waves (electromagnetic energy radiated per unit time) from a source, which makes the source appear luminous to the human eye. Unit used is the lumen. Lumen is a measure for light flux that incorporates the response of the typical human eye into its definition and, luminous efficacy of a source is a measure of how well the source is able to make itself appear luminous to the human eye. If a source radiates 1 Watt as monochromatic radiation of wavelength $540 \times 10^{12}$ Hz (to which the human eye is most sensitive) its luminous efficacy is taken as 683 lumen / Watt. Candela is the unit of luminous intensity– luminous flux emitted into unit solid angle in a given direction. A common wax candle emits light with a luminous intensity of roughly one candela.

Box 2. The SI prefixes

| Factor | Name | Symbol | Factor | Name | Symbol |
|---|---|---|---|---|---|
| $10^1$ | deca | da | $10^{-1}$ | deci | d |
| $10^2$ | hecto | h | $10^{-2}$ | centi | c |
| $10^3$ | kilo | k | $10^{-3}$ | milli | m |
| $10^6$ | mega | M | $10^{-6}$ | micro | μ |
| $10^9$ | giga | G | $10^{-9}$ | nano | n |
| $10^{12}$ | tera | T | $10^{-12}$ | pico | p |
| $10^{15}$ | peta | P | $10^{-15}$ | femto | f |
| $10^{18}$ | exa | E | $10^{-18}$ | atto | a |
| $10^{21}$ | zetta | Z | $10^{-21}$ | zepto | z |
| $10^{24}$ | yotta | Y | $10^{-24}$ | yocto | y |

[Table courtesy: Ref [1]]

Box 3. Chronology – Key steps in the history of the International System of Units (SI)
*(Excerpted from the press kit of the 26th General Conference on Weights and Measures – 2018, published by the BIPM)*

*17 April 1795*: The Metric System established in France by law.

*22 June 1799*: Two platinum standards representing the metre and the kilogram were deposited in the French National Archives.

*1832*: Carl Friedrich Gauss introduced a system of "absolute" units based on the millimetre, the milligram and the second.

*1st Sept 1869*: Emperor Napoleon III approved the creation of an international scientific commission to propagate the use of metric measurement to facilitate trade, the comparison of measurement between states and the creation of an international metre prototype.

*16th Nov1869*: The French government invited countries to join the International Scientific Commission.

*1870*: The first meeting of the newly formed International Metre Commission.

*1874*: The British Association for the Advancement of Science introduced the CGS (centimetre, gram and second) System.

*20 May 1875*: The signing of the Metre Convention on 20th May, by 17 countries, established The General Conference on Weights and Measures (CGPM) to discuss and endorse proposed changes to the system of units and The International Committee for Weights and Measures (CIPM), to oversee the discussion and recommendations on the system of units and the establishment of The International Bureau of Weights and Measures (BIPM) to provide the administration of the entire system and to house the international prototypes standards. The system of units agreed was similar to the CGS but called MKS with base units of the metre, kilogram and second. The metre and kilogram were represented by physical artefacts and the second by the astronomical second.

*1901*: Giovanni Giorgi proposed to the Associazione Elettrotecnica Italiana a new system enabling the combination of the fundamental units, the kilogram, the metre and the second, with a fourth unit of an electrical nature.
*1927*: The Consultative Committee for Electricity (CCE, now the Consultative Committee for Electricity and Magnetism CEPM) was created by the CIPM. It was the first such committee.
*1948*: The 9th CGPM requested the CIPM to launch an international survey, the outcome of which was to be used to formulate recommendations for a single practical system of measurement units, suitable for adoption by all countries.
*1954*: The CGPM approved the introduction of the ampere, the kelvin and the candela as base units for electric current, thermodynamic temperature and luminous intensity respectively.
*1960*: The 11th CGPM adopted the name of the International System of Units (SI) for the system based on six base units: the metre, the kilogram, the second, the ampere, the kelvin and the candela.
*1967*: The second was redefined as an "atomic second". The new definition depended henceforth on the properties of a caesium atom.
*1971*: The 14th CGPM added a new unit to the SI: the mole as the unit for amount of substance.
*1979*: The candela was redefined in terms of a monochromatic radiation.
*1983*: For the first time a definition of a base unit of the SI was based on a fundamental constant: the speed of light. The metre was henceforth the length of the path travelled by light in vacuum during a specific fraction of a second.
*1990*: New practical conventions based on quantum phenomena were adopted for the ohm and the volt.
*16 Nov 2018*: Four base units of the SI will be redefined: each definition will be linked to a constant of physics.
*20 May 2019*: World Metrology Day and official entry into force of the revised SI as the culmination of many years of intensive scientific cooperation between the National Metrology Institutes and the BIPM.

### Box 4. Electrical units and the Josephson and von Klitzing constants

The track is as follows [5]. The base unit of electricity, the ampere, was defined in terms of the force per unit length of two idealized, parallel wires carrying identical current; details of this definition fix the value of $\mu_0$, the magnetic constant, to be $4\pi \times 10^{-7}$ N/A$^2$ exactly. Now, the electric constant $\varepsilon_0$ is fixed by the relation $c^2\mu_0\varepsilon_0 = 1$ using the value fixed for $c$ the speed of light in vacuum. Then electrical metrologists ran into a conundrum as follows. A) It is known that a) given an accurately known microwave frequency $f$ (which is countable) d.c. voltages can be related to the Josephson constant $K_J$ (which is equal to $2e/h$ as per the Josephson effect relation[2]) via $U = nK_J^{-1} f$, where $U$ is the voltage measured and $n$ is an integer selected by the metrologist b) known resistances $R_H$ could be related to the von Klitzing constant $R_K$ (which is equal to $h/e^2$ as per the integer quantum Hall effect relation[3]) via $R_H = R_K / i$ where $i$ is an integer, selected by the metrologist. B) As far as measurements are concerned, in electrical metrology, a) voltages and resistances measured accurately using a variety of Josephson and Hall devices (quantum electrical devices) can be compared to very high precision and they appear identical b) the dimensionless fine-structure constant $\alpha = \mu_0 c e^2 / 2h$ can also be measured with relative uncertainty less than $1 \times 10^{-9}$. C) Now, the fundamental unit of electric charge $e$ can be derived from the measured value of $\alpha$, using a measured value for $h$. D) But, a) The measurement of $h$ (dimension ML$^2$T$^{-1}$) is tied to the IPK, thus tying the measured value of $e$ to the IPK b) the measurement of $h$ is difficult and $h$ is measured with an uncertainty that is more than the uncertainties in the various quantum electrical standards. Now comes the conundrum. The relative imprecision of $h$ spoils the precision of electrical standards whose determination uses the value of $h$ and, every time a new value is recommended for $h$, all quantities, determined in relation to the value of $h$ will have to be re-evaluated!

*Disconnecting the electrical units from the IPK*: The solution agreed upon was to fix the value of $h$ and that of $e$, make $\mu_0$ a quantity derived from the measured value of the fine structure constant and, fix the ampere using $e$ and the new second. This would disconnect the electrical units from the IPK. But, $h$ had to be determined to a level of accuracy matching that of the electrical measurements first, which is where the crystal balls helped.

Margin Note 2

In the d.c. Josephson effect an alternating supercurrent of frequency $2eV/h$ appears across two weakly coupled superconductors maintained at a potential difference $V$. In the a.c. Josephson effect, when a RF field is applied across a weakly coupled junction the dc I-V characteristics show steps at bias voltages which are integer multiples of $hf/2e$, where f is the frequency of the RF field applied – for more on this see [6].

Margin Note 3

In the quantum Hall effect electrons are confined to move in a plane. This is achieved by what is called an inversion layer which is an arrangement of semiconductors. The layer is kept at a very low temperature and a large magnetic field is applied perpendicular to the plane. With this arrangement, on measuring the resistance, von Klitzing found that the Hall resistance, equal to the in-plane voltage developed across the sample ÷ current flowing perpendicular to the plane, showed well defined plateaus as a function of the magnetic field. The values of the Hall resistance were integer multiples of $h/e^2$ – for more on this see [7].

Box 5. Mass measurement with the watt balance

In the watt balance (Figure 4a) measurements in the weighing and moving modes (refer Fig 4b) finally yield the relation $mgv = U_m I_w$ which is like comparing power and hence the name watt balance. In the weighing mode the Lorentz force just balances the weight of the mass to be measured, yielding the relation $mg = BLI_w$. In the moving mode, Faraday's law of induction yields the relation $U_m = BLv$. Eliminating $BL$ between the two relations gives the mass as $m = U_m I_w / gv$.

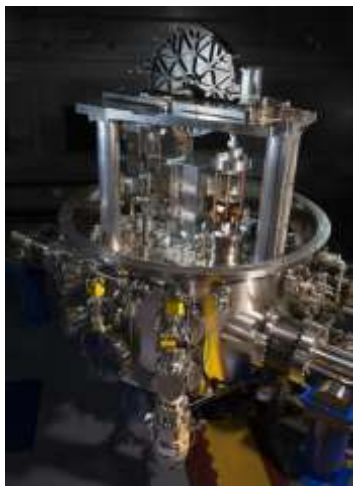
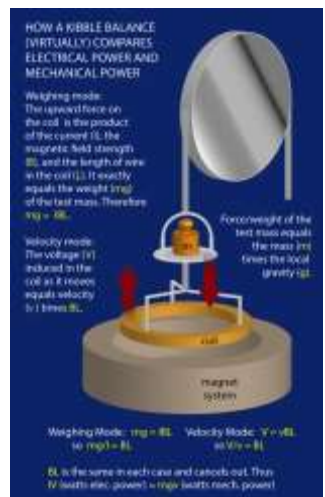

Figure 4a: A watt balance  Figure 4b: The principle of the watt balance
(*Figure courtesy:* J. L. Lee / NIST)  (*Figure courtesy*: Suplee/NIST)

When in operation the whole setup is covered by a vacuum enclosure to avoid buoyancy effects. Currents and voltages are precisely measured using quantum electrical effects – voltages by making use of the Josephson effect and resistances by making use of the integer quantum Hall effect. The gravity at the place is measured using an absolute gravimeter; motion of the coil is tracked by laser interferometry, time intervals are determined using an atomic clock.

Box 6. Disconnecting the kilogram from the IPK – the details

Based on the equation given below, the mass of a well characterized, homogeneous, perfect, $^{28}$Si single crystal sphere can be determined as a function of the Planck constant, without reference to any other mass standard; thus making the silicon sphere a primary mass standard.

$$m_{artefact} = \frac{2hR_\infty}{c\alpha^2} \frac{\sum_i x(^iSi)A_r(^iSi)}{A_r(e)} \frac{8V_{core}}{a^3} - m_{deficit} + m_{SL}$$

In this equation $2hR_\infty/c\alpha^2$ is the mass of the electron, $\sum_i x(^iSi)A_r(^iSi)/A_r(e)$ is the mean mass ratio of silicon to the mass ratio of the electron (the index $i = 28, 29, 30$ running over the atomic masses of the various isotopes of Si), $8V_{core}/a^3$ is the number of silicon atoms in the core of the crystal ball, $m_{deficit}$ is the influence of point defects (i.e. impurities and self-point defects in the crystal) on the core mass and $m_{SL}$ is the mass of the surface layer. The terms may be unpacked further as follows: $R_\infty$ – Rydberg constant, $\alpha$ – fine structure constant; $x(^iSi)$ – the amount-of-substance fraction of isotope $^iSi$ in the crystal, $A_r(^iSi) = 12m_{^iSi}/m_{^{12}C}$ – the relative atomic mass of each isotope $^iSi$ with respect to the atomic mass unit (relative masses may be determined with much less uncertainty than the masses themselves), $A_r(e) = 12m_e/m_{^{12}C}$ is the relative mass of the electron; $V_{core}$ is the volume of the core of the silicon crystal ball excluding the volume of the surface layers which get contaminated by adsorption of water, by carbon impurities and by oxidation, $a$ – lattice constant of the silicon crystal. $m_{deficit} = V_{core} \sum_x (m_{28} - m_x)N_x$; $m_{28}$ and $m_x$ are the masses of a $^{28}$Si atom and the point defect referred to as $x$ (Carbon, Nitrogen, Boron etc.) respectively, $N_x$ is the concentration of the point defect $x$. For a vacancy, $m_x = m_{vacancy} = 0$. $m_{SL}$ is determined via two different sets of experimental techniques, one directly determining the mass of the surface layers (physisorbed water layer, chemisorbed water layer, carbonaceous contamination, oxide layer) and the other determining the thickness of the surface layer and obtaining the mass of the layer using an assumed density for the layer. Of the above, the atomic and electrical constants are known to high accuracy and $h$ and $c$ are fixed in the new SI. Highly defect free, high purity, high $^{28}$Si isotope fraction, highly perfectly shaped one kilogram spheres cut from specially grown silicon crystals are well characterized to yield the remaining factors on the RHS. A flavour of the complexities involved in attaining this particular feat of metrology may be obtained from noticing that a paper from 2016 detailing the then status of the 'Realization of the kilogram by the XRCD method' [9] mentions altogether, more than fifty different highly sophisticated experimental techniques / processes for 'Isotope enrichment, crystal production, and manufacturing of silicon spheres', 'Crystal perfection: evaluation of point defects', 'Lattice parameter measurement', 'Molar mass measurement', 'Surface evaluation for silicon spheres', 'Diameter and volume measurements' and 'Mass measurement'.